\documentclass[12pt]{article}
\usepackage{epsfig}
\usepackage{amssymb}
\usepackage{amsmath}
\usepackage{amsfonts}

\newcommand{\R}{\mathbb{R}}
\newcommand{\C}{\mathbb{C}}
\newcommand{\Z}{\mathbb{Z}}

\newcommand{\be}{\begin{equation}}
\newcommand{\ee}{\end{equation}}
\newcommand{\bea}{\begin{eqnarray}}
\newcommand{\eea}{\end{eqnarray}}
\newcommand{\nn}{\nonumber}
\newcommand{\kt}{\rangle}
\newcommand{\br}{\langle}
\newcommand{\cun}{\mbox{\scriptsize${\cal N}$}}

\newcommand{\ed}{\end{document}}

\newcommand{\rx}{{\rm x}}

\newcommand{\rp}{{\rm p}}

\begin{document}

\title{Pseudo-Hermitian Description of $PT$-Symmetric
Systems Defined on a Complex Contour}

\author{\\
Ali Mostafazadeh
\\
\\
Department of Mathematics, Ko\c{c} University,\\
34450 Sariyer, Istanbul, Turkey\\ amostafazadeh@ku.edu.tr}
\date{ }
\maketitle

\begin{abstract}
We describe a method that allows for a practical application of
the theory of pseudo-Hermitian operators to $PT$-symmetric systems
defined on a complex contour. We apply this method to study the
Hamiltonians $H=p^2+x^2(ix)^\nu$ with $\nu\in(-2,\infty)$ that are
defined along the corresponding anti-Stokes lines. In particular,
we reveal the intrinsic non-Hermiticity of $H$ for the cases that
$\nu$ is an even integer, so that $H=p^2\pm x^{2+\nu}$, and give a
proof of the discreteness of the spectrum of $H$ for all
$\nu\in(-2,\infty)$. Furthermore, we study the consequences of
defining a square-well Hamiltonian on a wedge-shaped complex
contour. This yields a $PT$-symmetric system with a finite number
of real eigenvalues. We present a comprehensive analysis of this
system within the framework of pseudo-Hermitian quantum mechanics.
We also outline a direct pseudo-Hermitian treatment of
$PT$-symmetric systems defined on a complex contour which
clarifies the underlying mathematical structure of the formulation
of $PT$-symmetric quantum mechanics based on the
charge-conjugation operator. Our results provide a conclusive
evidence that pseudo-Hermitian quantum mechanics provides a
complete description of general $PT$-symmetric systems regardless
of whether they are defined along the real line or a complex
contour.
\end{abstract}
\begin{center}
~~~~~PACS number: 03.65.-w
\end{center}



\newpage

\section{Introduction}

The notion of a pseudo-Hermitian operator as outlined in
\cite{p1,p2,p3} provides a general framework for understanding the
intriguing mathematical properties of $PT$-symmetric Hamiltonians
\cite{bender-98,bender-99}.\footnote{The term ``pseudo-Hermitian''
has been in use within the context of indefinite-metric quantum
theories \cite{indefinite-phys} and indefinite-metric linear
spaces \cite{indefinite-math} since the 1940's, \cite{mielnik}. In
this context it corresponds to what is termed as
``$\eta$-pseudo-Hermitian'' in \cite{p1}, where $\eta$ is an {\em
a priori} fixed indefinite metric operator. The relevance of the
indefinite-metric theories and $PT$-symmetric systems has been
considered in \cite{Japaridze}. The definition of a
pseudo-Hermitian operator given in \cite{p1} (and used below) is
slightly different from the one used in earlier publications,
e.g., \cite{indefinite-phys,indefinite-math,Japaridze}. As
explained in detail in \cite{cjp-2003}, this slight difference has
important conceptual and technical ramifications. In particular,
together with the idea of using biorthonormal systems \cite{p1} it
opens up the way for the construction of all possible metric
operators, leads to the important observation that there is a
positive-definite inner product rendering the Hamiltonian
Hermitian for the cases that the spectrum is real \cite{p2}, and
reveals the nature of the connection with antilinear symmetries
such as $PT$, \cite{p3}.} It involves an underlying Hilbert space
${\cal H}$ in which the operator acts. For $PT$-symmetric
Hamiltonians defined on the real line, ${\cal H}$ is the familiar
space of square-integrable functions. For the $PT$-symmetric
Hamiltonians $H$ defined on a complex contour and having a
discrete spectrum, ${\cal H}$ is the Hilbert space obtained by
Cauchy-completing the span of the eigenfunctions of $H$ with
respect to an arbitrarily chosen positive-definite inner product
\cite{jpa-2003,cjp-2003,cjp-2004b}. The implicit nature of this
construction makes a direct application of the theory of
pseudo-Hermitian operators for these Hamiltonians intractable.
This forms the basis of the view that this theory is incapable of
dealing with $PT$-symmetric Hamiltonians defined on a complex
contour. The purpose of this article is to show that indeed the
opposite is true. This is done by an explicit construction that
allows for the description of the same system using the
information given on the real axis. It reveals the implicit
non-Hermiticity of the apparently Hermitian $PT$-symmetric
Hamiltonians, such as $p^2-x^{4}$, that are defined along an
appropriate complex contour \cite{bender-98,bender-99}.
Furthermore, it leads to a previously unnoticed connection between
the spectral properties of the $PT$-symmetric Hamiltonians of the
form
    \[H=p^2+x^2(ix)^\nu,\]
(defined on an appropriate contour) and those of the Hamiltonians
of the form
    \[H=p^2+|x|^{2+\nu},\]
(which are obtained by requiring the eigenfunctions to belong to
$L^2(\R)$ and satisfy certain boundary conditions at $x=0$.) An
important advantage of a pseudo-Hermitian description of
$PT$-symmetric systems defined on a complex contour is that it
offers a prescription for computing the physical observables
\cite{cjp-2004b,p61,p64} of these theories.

In the remainder of this section we include a brief review of the
relevant aspects of the theory of pseudo-Hermitian operators. For
clarity of the presentation we will only consider Hamiltonian
operators that have a discrete nondegenerate spectrum. In
particular, we will focus our attention mainly on the cases that
the spectrum is not only discrete and nondegenerate but also real
(and bounded from below). It is an operator with the latter
properties that can serve as the Hamiltonian for a unitary quantum
system, \cite{jmp-2004}. If complex eigenvalues are present, we
identify the vector space underlying the physical Hilbert space
with the span of the eigenfunctions with real eigenvalues and
restrict the Hamiltonian to this vector space
\cite{jpa-2003,cjp-2003,cjp-2004b}.

Let ${\cal H}$ be a given separable Hilbert space with inner
product $\br\cdot|\cdot\kt$ and $H:{\cal H}\to{\cal H}$ be a
linear operator. Then $H$ is called a pseudo-Hermitian operator
\cite{p1} if there exists a Hermitian invertible operator
$\eta:{\cal H}\to{\cal H}$ satisfying
    \be
    H^\dagger=\eta H\eta^{-1},
    \label{ph}
    \ee
where for any linear operator $A:{\cal H}\to{\cal H}$, $A^\dagger$
stands for the `adjoint of $A$', i.e., the unique operator
$A^\dagger:{\cal H}\to{\cal H}$ satisfying $\br\cdot|A\cdot\kt=
\br A^\dagger\cdot|\cdot\kt$. The operator $\eta$ entering the
defining relation~(\ref{ph}), which is sometimes referred to as a
metric operator, is not unique \cite{p4,jmp-2003}. In fact the set
${\cal U}_H$ consisting of all metric operators is always an
infinite set. A simple property of a pseudo-Hermitian operator is
that it is Hermitian with respect to the possibly indefinite inner
product $\br\cdot,\cdot\kt_\eta:=\br\cdot|\eta\cdot\kt$, i.e.,
$\br\cdot, H\cdot\kt_\eta= \br H\cdot,\cdot\kt_\eta$, \cite{p1}.

Next, suppose that $H$ has a complete set of eigenvectors
$\psi_n\in{\cal H}$, i.e., it is diagonalizable. Then one can
construct the vectors $\phi_n\in{\cal H}$ that together with
$\psi_n$ form a biorthonormal system for the Hilbert space, i.e.,
    \be
    \br\phi_n|\psi_m\kt=\delta_{mn},~~~~~~~~
    \sum_n|\psi_n\kt\br\phi_n|=1.
    \label{biortho}
    \ee
Using the properties of such biorthonormal systems, one can prove
the following characterization theorem \cite{p2}.
    \begin{itemize}
    \item[] {\bf Theorem:} For a diagonalizable linear operator $H$
    with a discrete spectrum the following conditions are
    equivalent.
        \begin{itemize}
        \item[(c1)] The spectrum of $H$ is real.
        \item[(c2)] $H$ is pseudo-Hermitian and the set ${\cal U}_H$
        includes a positive-definite metric operator $\eta_+$.
        \item[(c3)] $H$ is Hermitian with respect to a
        positive-definite inner product $\br\cdot,\cdot\kt_+$,
        e.g., $\br\cdot,\cdot\kt_{\eta_+}:=\br\cdot|\eta_+\cdot\kt$.
        \item[(c4)] $H$ may be mapped to a Hermitian operator
        $h:{\cal H}\to{\cal H}$ via a similarity transformation,
        i.e., there is an invertible operator $\rho:{\cal
        H}\to{\cal H}$ such that
            \be
            h:=\rho H\rho^{-1}
            \label{hermitian-h}
            \ee
        is Hermitian.
        \end{itemize}
        \end{itemize}
If one (and therefore all of) these conditions hold, one has the
following spectral resolutions for $H$ and $H^\dagger$.
    \be
    H=\sum_n E_n|\psi_n\kt\br\phi_n|,~~~~~~~~~
    H^\dagger=\sum_n E_n|\phi_n\kt\br\psi_n|.
    \label{resol}
    \ee
Furthermore, a positive-definite metric operator $\eta_+$ is given
by
    \be
    \eta_+=\sum_n|\phi_n\kt\br\phi_n|,
    \label{eta+}
    \ee
and a canonical example of the invertible operator $\rho$ whose
existence is guaranteed by condition (c4) is
$\rho=\sqrt{\eta_+}$.\footnote{For a mathematically rigorous
discussion of pseudo-Hermitian operators, see
\cite{albeverio-kuzhel}.}

The metric operator $\eta_+$ plays the same role in
pseudo-Hermitian quantum mechanics \cite{cjp-2003} as the metric
tensor does in general relativity, \cite{pla-2004}. It allows for
the construction of the physical Hilbert space ${\cal H}_{\rm
phys}$ and the observables of the system. The Hilbert space ${\cal
H}_{\rm phys}$ has the same vector space structure as ${\cal H}$
but its inner product is given by
    \be
    \br\cdot,\cdot\kt_+=\br\cdot,\cdot\kt_{\eta_+}
    :=\br\cdot|\eta_+\cdot\kt.
    \label{inn-pr}
    \ee
The observables $O$ of the theory are linear Hermitian operators
acting in ${\cal H}_{\rm phys}$, \cite{p61}. They can be obtained
from the Hermitian operators $o$ acting in ${\cal H}$ according to
    \be
    O=\rho^{-1}o\rho.
    \label{observable}
    \ee
The formulation of the dynamics and the interpretation of the
theory are identical with those of the conventional quantum
mechanics. {\em Pseudo-Hermitian quantum mechanics shares all the
postulates of conventional quantum mechanics except that the inner
product of the physical Hilbert space ${\cal H}_{\rm phys}$ is not
a priori fixed but determined by the eigenvalue problem for a
linear (Hamiltonian) operator that acts on a reference Hilbert
space ${\cal H}$.}

As we mentioned above the formulation of the theory does not fix
the reference Hilbert space ${\cal H}$. For systems with a
finite-dimensional state space, one usually identifies ${\cal H}$
with the complex Euclidean space, i.e., $\C^N$ with usual
Euclidean inner product: $\br \vec \psi|\vec
\phi\kt:=\vec\psi^*\cdot\vec\phi$, where a dot means ordinary dot
product of vectors, \cite{jpa-2003}. For $PT$-symmetric theories
defined on the real axis, e.g., for $H=p^2+ix^3$, the natural
choice for ${\cal H}$ is $L^2(\R)$, \cite{jmp-2003}. However, for
$PT$-symmetric theories that are defined on a complex contour
$\Gamma$, such as $H=p^2-x^4$, a natural and useful choice for the
reference Hilbert space ${\cal H}$ has not been available. The
main purpose of this article is to offer a satisfactory resolution
of this problem by showing how one can formulate and describe the
same theories using equivalent $PT$-symmetric Hamiltonians whose
eigenvalue problem is defined in $L^2(\R)$. This `real
description' facilitates the understanding of the physical content
of these theories. It allows us to use the usual mathematical
tools of conventional quantum mechanics and deal with the
manifestly non-Hermitian form of the Hamiltonians such as
$H=p^2-x^4$ whose non-Hermiticity stems from their domain of
definition rather than their explicit form. An alternative but
less practical approach is to develop a pseudo-Hermitian
description of $PT$-symmetric systems that is based on the choice:
${\cal H}=L^2(\Gamma)$. This `complex description' clarifies the
underlying mathematical structure of the formulation of
$PT$-symmetric quantum mechanics that is based on the
charge-conjugation operator \cite{bbj,Weigert,pla-bj}.

\section{Moving Back to the Real Line}

Suppose ${\cal F}$ is the set of real-analytic
functions\footnote{Notice that ${\cal F}\cap L^2(\R)$ is a dense
subset of $L^2(\R)$.} $\psi:\R\to\C$ and $H:{\cal F}\to{\cal F}$
is a linear operator of the form
    \be
    H=[  p-A(  x)]^2+V(  x),
    \label{H}
    \ee
where $A,V:\R\to\C$ are piecewise real-analytic functions,
$p\psi(x):= -i\psi'(x)$ for all $\psi\in{\cal F}$, and a prime
stands for a derivative. A particularly well-studied example is
    \be
    H=p^2+x^2(ix)^\nu,~~~~~~~\nu\in(-2,\infty).
    \label{H=}
    \ee
The main observation that has led to the current interest in
$PT$-symmetric quantum mechanics is that for certain non-real
choices of $V$ (and $A=0$), for example (\ref{H=}) with $\nu\geq
0$, the operator $H$ has a real and discrete spectrum provided
that its eigenvalue problem is solved along an appropriate contour
$\Gamma$ in the complex plane, \cite{bender-98}.\footnote{A
mathematically rigorous proof of this statement is given in
\cite{dorey,shin}.} This was a rather intriguing observation
because generically the operator $H$, which we will call the
Hamiltonian, is manifestly non-Hermitian with respect to the
$L^2$-inner product.

A typical physicist who is not familiar with the subject would
immediately reject the statement that ``$H=p^2-x^4$ has a discrete
spectrum.''\footnote{ This Hamiltonian corresponds to the choice
$\nu=2$ in (\ref{H=}).} Indeed, This statement is neither true nor
false, because the eigenvalue problem for a linear operator
defined on an infinite-dimensional vector space is well-posed only
for specific choices of the domain of the operator. In the case of
differential operators such as (\ref{H}), in particular
(\ref{H=}), the determination of the domain is related to the
choice of the asymptotic boundary conditions. A nontrivial
observation made in \cite{bender-98} is that one obtains a
discrete spectrum for (\ref{H=}) provided that one imposes the
asymptotic boundary conditions along an appropriate contour
$\Gamma$ in the complex plane.\footnote{For the cases that $\nu$
is an integer greater than $-2$, so that the potential term in
(\ref{H=}) is a monomial, this was known to mathematicians
\cite{sibuya}. We will give a proof of the discreteness of the
spectrum for all $\nu\in(-2,\infty)$ in the Appendix.} This means
that one has to identify the eigenvalue equation for (\ref{H})
with its complex (holomorphic) extension \cite{Hille,sibuya}
    \be
    \left\{-\left[\frac{d}{dz}-iA(z)\right]^2+V(z)\right\}
    \Psi_n(z)=E_n\Psi_n(z),
    \label{eg-va}
    \ee
and seek for solutions $\Psi_n$ such that
    \be
    |\Psi_n(z)|\to 0~\mbox{exponentially as $z$ moves off to
    the infinity along $\Gamma$}.
    \label{bound}
    \ee
Note that the contour $\Gamma$ is generally the graph of a
(continuous piecewise) regular curve \cite{diff-geo} parameterized
by $s\in\R$, i.e., there is a (continuous piecewise)
differentiable function $\zeta:\R\to\C$ with non-vanishing first
derivative such that
    \be
    \Gamma=\{\zeta(s)|s\in\R\},
    \label{gamma}
    \ee
and that $\lim_{s\to\pm\infty} \Re[\zeta(s)]=\pm\infty$. Here and
in what follows $\Re$ and $\Im$ respectively mean `real' and
`imaginary part of'. Clearly, we may state the boundary condition
(\ref{bound}) as
    \be
    |\Psi_n(\zeta(s))|\to 0~\mbox{exponentially as $s\to\pm\infty$}.
    \label{bound-2}
    \ee
For the Hamiltonians~(\ref{H=}), it is the choice of an
appropriate contour $\Gamma$ and the imposition of the boundary
conditions (\ref{bound}) that lead to a discrete set of nontrivial
solutions for (\ref{eg-va}). The same holds for various
generalizations of (\ref{H=}), \cite{bender-99,shin-2004}. In
general the contour $\Gamma$ is not uniquely determined by the
mathematical considerations, though it is required to stay in the
so-called Stokes wedges in the asymptotic region, i.e., where
$s\to\pm\infty$. In particular, there is a preferred choice for
the asymptotic shape of $\Gamma$ that maximizes the decay rate of
the solutions of (\ref{eg-va}). This corresponds to the bisector
of the appropriate Stokes wedge. Making this choice for the
Hamiltonians (\ref{H=}) we have \cite{bender-98}
    \be
    \lim_{s\to\pm\infty}{\rm arg}[\zeta(s)]=-
    \theta^\pm_\nu,
    \label{stokes}
    \ee
where `arg' abbreviates `argument of' and
    \be
    \theta^+_\nu=\theta_\nu:=\frac{\pi\,\nu}{2(\nu+4)},~~~~~~~
    \theta^-_\nu:=\pi-\theta_\nu.
    \label{stokes2}
    \ee

Next, we identify the real and imaginary axes of $\C$ with the
$x$- and $y$-axes of the usual Cartesian coordinate system on
$\R^2=\C$, so that $z=x+iy$, and consider a general smooth contour
$\Gamma$ such that $\Re[\Gamma(x+iy)]$ is an increasing function
of $x:=\Re(z)$.\footnote{This is not a strong condition. One can
always choose such a contour for the purpose of defining boundary
conditions (\ref{bound}).} Then we can express the function
$\zeta$ in terms of a differentiable real-valued function
$f:\R\to\R$ according to
    \be
    \zeta(x)=x+if(x).
    \label{zeta}
    \ee
The condition that $\zeta$ is a regular curve is also satisfied,
because $|\zeta'(x)|^2=1+f'(x)^2\neq 0$.

Now, we wish to restrict the complex differential equation
(\ref{eg-va}) to the contour $\Gamma$, and obtain an equivalent
real differential equation with generally complex coefficients.
Along $\Gamma$ we have $z=\zeta(x)=x+if(x)$. A simple change of
variable $z\to x+i f(x)$ in (\ref{eg-va}) yields
    \be
    \left\{-g(x)^2\left[\frac{d}{dx}-i a(x)\right]^2+
    ig(x)^3f''(x)\left[\frac{d}{dx}-i a(x)\right]+ v(x)\right\}
    \psi_n(x)=E_n\psi_n(x),
    \label{eg-va-r}
    \ee
where
    \bea
    g(x)&:=&[\zeta'(x)]^{-1}=[1+if'(x)]^{-1},~~~~~~
    a(x):=g(x)^{-1}A[x+if(x)],
    \label{new-1}\\
    v(x)&:=&V[x+if(x)],~~~~~~~~~~~~~\psi_n(x):=\Psi_n[x+if(x)].
    \label{new-2}
    \eea
The complex differential equation (\ref{eg-va}) together with the
boundary condition (\ref{bound}) (alternatively (\ref{bound-2}))
is clearly equivalent to real differential equation
(\ref{eg-va-r}) together with the boundary condition
    \be
    |\psi(x)|\to 0~~ \mbox{exponentially as}~~ |x|\to\infty.
    \label{bound-3}
    \ee
The analyticity properties \cite{Hille} of $\Psi_n$ and
consequently of $\psi_n$ together with the condition
(\ref{bound-3}) implies that $\psi_n\in L^2(\R)$. In other words,
the eigenvalue problem for the Hamiltonian~(\ref{H}) defined by
Eq.~(\ref{eg-va}) is equivalent to the eigenvalue problem for the
Hamiltonian
    \be
    H':=g(  x)^2[  p-a(  x)]^2-
    g(  x)^3f''(  x)[  p-a(  x)]+v(  x).
    \label{H-prime}
    \ee
viewed as an operator acting in $L^2(\R)$.

\section{Consequences of Imposing $PT$-symmetry}

Let $\xi:\R\to\C$ be a function. Then under the joint action of
the parity $P$ and time-reversal $T$ operators, $\xi(  x)\to
PT\,\xi( x)\,PT=\xi(-  x)^*$. Applying this rule to the
Hamiltonian (\ref{H-prime}) and using $PT\,   p\, PT=  p$, we find
    \be
    PT\, H'\, PT=g(-  x)^{*2}[  p-a(-  x)^*]^2-
    g(-  x)^{*3}f''(-  x)^*[  p-a(-  x)^*]+v(-  x)^*.
    \label{H-prime-PT}
    \ee
In particular, demanding $H'$ to be PT-symmetric yields
    \bea
    g(-  x)^{*2}&=&g(  x)^{2},~~~~~~
    g(-  x)^*f''(-  x)^*=g(-  x)^*f''(  x),
    \label{x1}\\
    a(-  x)^*&=&a(  x),~~~~~~~
    v(-  x)^*=v(  x).
    \label{x2}
    \eea
In view of Eqs.~(\ref{new-1}), (\ref{new-2}), (\ref{x1}), and
(\ref{x2}), the fact that $f$ is a real-valued function, and $x$
takes zero as a value, we have
    \be
    f(x)=f(-x),~~~~~~\left. A(u)^*\right|_{u=-[x+if(x)]}=
    A[x+if(x)],~~~~~~\left. V(u)^*\right|_{u=-[x+if(x)]}=
    V[x+if(x)].
    \label{PT}
    \ee
The first of these equations imply that along the contour
$\Gamma$, $z(-x)^*=-z(x)$. Therefore the condition that $H'$ be
$PT$-symmetric implies that $\Gamma$ has reflection-symmetry about
the $y$- (or imaginary-) axis. The second and third equations in
(\ref{PT}) and the assumption that $A$ and $V$ may be analytically
continued onto the contour $\Gamma$ indicate that they are
separately $PT$-symmetric, i.e.,
    \be
    PT\, A(  x)\, PT=A(  x),~~~~~~~~~~
    PT\, V(  x)\, PT=V(  x).
    \label{PT-2}
    \ee
These are equivalent to requirement that the original Hamiltonian
(\ref{H}) be $PT$-symmetric.

In summary, the Hamiltonian (\ref{H}) and the contour $\Gamma$ are
$PT$-symmetric if and only if the Hamiltonian (\ref{H-prime}) is
$PT$-symmetric. In the following we will only consider the cases
that these conditions hold.

\section{Wedge-Shaped Contours}

\begin{figure}[t]
    \centerline{\epsffile{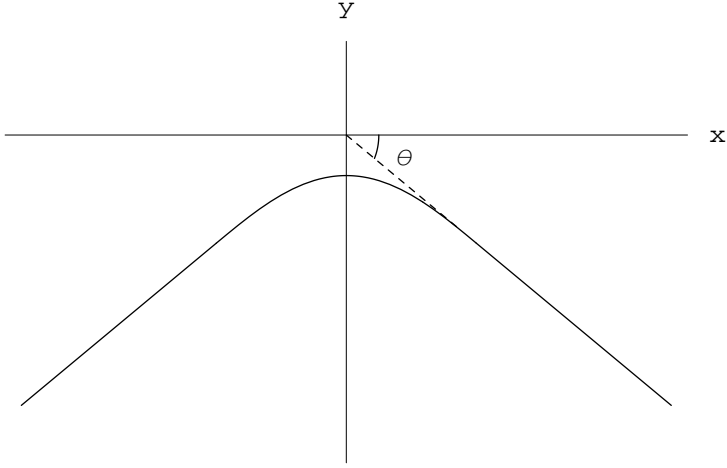}}
    \centerline{\parbox{11cm}{\caption{Plot of
    $y=f_\epsilon(x)$. $f_\epsilon$ has a maximum at $x=0$ with value
    $f_\epsilon(0)=-3\epsilon\tan\theta/8$. The angle $\theta$ is
    also displayed} \label{fig1}}}
    \end{figure}

The simplest possible $PT$-symmetric choices for the contour
$\Gamma$ are the wedge-shaped contours:
    \be
    \Gamma(x)= x[1-i\, {\rm sign}(x)\tan \theta],
    \label{lambda-shaped}
    \ee
where ${\rm sign}(x):=x/|x|$ for $x\neq 0$, ${\rm sign}(0):=0$,
and $\theta\in[0,\pi/2)$. Clearly $\Gamma$ is not a regular curve
at $x=0$. Therefore, we will smoothen it in a small neighborhood
of $x=0$, say according to $\Gamma\to\Gamma_\epsilon$, where
    \bea
    \Gamma_\epsilon(x)&:=&x+i f_\epsilon(x),
    \label{smooth-G}\\
    f_\epsilon(x)&:=&\left\{\begin{array}{ccc}
    -|x|\tan\theta & {\rm for} &|x|\geq \epsilon\\
    \varphi_\epsilon(x)&
    {\rm for} &
    |x|\leq \epsilon,\end{array}\right.
    \label{smooth}\\
    \varphi_\epsilon(x)&:=&\frac{\epsilon\tan\theta}{8}\,
    \left[\left(\frac{x}{\epsilon}\right)^4
    -6\left(\frac{x}{\epsilon}\right)^2-3\right],
    \label{varphi}
    \eea
and $\epsilon\in\R^+$ is an arbitrary constant. Note that
$f_\epsilon$ is a twice-differentiable function that can be
substituted for $f$ in the expression~(\ref{H-prime}) for the
Hamiltonian $H'$ and that its maximum value is
$f_\epsilon(0)=-3\epsilon\tan\theta/8$. Figure~\ref{fig1} shows a
plot of $f_\epsilon$.
Furthermore, in view of (\ref{smooth-G}) and (\ref{smooth}), we
have
    \be
    \Gamma_\epsilon(x)=\left\{\begin{array}{ccc}
    \sec(\theta)\, e^{-i\theta\;{\rm sign}(x)}x & {\rm for} &|x|\geq \epsilon\\
    x+i\varphi_\epsilon(x)&{\rm for} &
    |x|\leq \epsilon.\end{array}\right.
    \label{G=}
    \ee

In what follows we shall consider the contours of the
form~(\ref{G=}) which yield the wedge-shaped
contours~(\ref{lambda-shaped}) in the limit $\epsilon\to 0$.

Setting $f=f_\epsilon$ in (\ref{new-1}) and using (\ref{smooth}),
we obtain
    \bea
    {\rm for}~~|x|\geq \epsilon:~&&~f'(x)=-\tan(\theta)\;{\rm
    sign}(x),~~~f''(x)=0,~~~
    g(x)=\cos(\theta)\, e^{i\theta\;{\rm sign}(x)},
    \label{new-5}\\
    {\rm for}~~|x|\leq \epsilon:~&&~f'(x)=\varphi'_\epsilon(x),~~~~
    f''(x)=\varphi''_\epsilon(x),~~~~g(x)=\gamma_\epsilon(x),
    \label{new-6}
    \eea
where
    \bea
    \varphi'_\epsilon(x)&:=&
    \frac{\tan\theta}{2}
    \left[\left(\frac{x}{\epsilon}\right)^3
    -3\left(\frac{x}{\epsilon}\right)\right],
    \label{phi-1}\\
    \varphi''_\epsilon(x)&:=&\frac{3\tan\theta}{2\epsilon}
    \left[\left(\frac{x}{\epsilon}\right)^2-1\right],
    \label{phi-2}\\
    \gamma_\epsilon(x)&:=&\left\{1+
    \frac{i\tan\theta}{2}
    \left[\left(\frac{x}{\epsilon}\right)^3
    -3\left(\frac{x}{\epsilon}\right)\right]\right\}^{-1}.
    \label{gamma2}
    \eea
These relations together with (\ref{new-1}), (\ref{new-2}), and
(\ref{H-prime}) then yield
    \be
    H'=H^{(\epsilon)}_-+H'_\epsilon+H^{(\epsilon)}_+,~~~~~~~~~~~
    H^{(\epsilon)}_\pm := \Lambda^{(\epsilon)}_\pm
    H_\pm \Lambda^{(\epsilon)}_\pm,~~~~~~~~~~
    H'_\epsilon:= \Lambda_\epsilon H_\epsilon \Lambda_\epsilon,
    \label{H-prime-2}
    \ee
where
    \bea
    \Lambda^{(\epsilon)}_+&:=&
    \int_\epsilon^\infty dx\: |x\kt\br x|,~~~~~
    \Lambda^{(\epsilon)}_-:=
    \int_{-\infty}^{-\epsilon}dx\: |x\kt\br x|,~~~~~
    \Lambda_\epsilon:=\int_{-\epsilon}^{\epsilon}dx\: |x\kt\br x|,
    \label{proj}\\
    H_\pm &:=& \cos^2(\theta)\, e^{\pm 2i\theta}\left\{  p-
    \sec(\theta)e^{\mp i\theta}A[\sec(\theta)
    e^{\mp i\theta}  x]\right\}^2+V[\sec(\theta)
    e^{\mp i\theta}  x],
    \label{H-pm}\\
    H_\epsilon &:=& \gamma_\epsilon(  x)^2[  p-a_\epsilon(  x)]^2-
    \gamma_\epsilon(  x)^3\varphi''_\epsilon(  x)
    [  p-a_\epsilon(  x)]+v_\epsilon(  x),
    \label{H-ep}\\
    a_\epsilon(x)&:=&\gamma_\epsilon(x)^{-1}
    A[x+i\varphi_\epsilon(x)],~~~~~~~
    v_\epsilon(x):=V[x+i\varphi_\epsilon(x)].
    \label{A-V}
    \eea

Note that $PT \Lambda^{(\epsilon)}_+ PT =\Lambda^{(\epsilon)}_-$
and $PT \Lambda_\epsilon PT =\Lambda_\epsilon$. These together
with (\ref{PT-2}) and (\ref{varphi}) -- (\ref{A-V}) yield the
following relations that are clearly consistent with the
$PT$-symmetry of $H'$.
    \be
    PT\: H^{(\epsilon)}_+\: PT =H^{(\epsilon)}_-,~~~~~~~~~~
    PT\: H_\epsilon\: PT =H_\epsilon.
    \label{PT-3}
    \ee

In practice, to solve the eigenvalue problem for $H'$, we may
solve the corresponding differential equation for
$|x|\geq\epsilon$ in the limit $\epsilon\to 0$ and match the
solution at $x=0$ by enforcing appropriate continuity
requirements. As we shall see below the latter yield a pair of
boundary conditions at $x=0$. It is the Hamiltonians $H_\pm$
together with these boundary conditions at $x=0$ and the
requirement: $\psi_n\in L^2(\R)$ that determine the eigenvalues
$E_n$.

The Hamiltonians $H_\pm$ take a simpler form in terms of the
scaled position and momentum operators:
    \be
     \rx:=\frac{  x}{\cos\theta},~~~~~~~~~~
     \rp:=\cos\theta\:  p.
    \label{scaled}
    \ee
The classical analog of $\rx$ corresponds to the arc-length
parametrization of the contour $\Gamma$, \cite{diff-geo}. Using
(\ref{H-pm}) and (\ref{scaled}), we have
    \be
    H_\pm :=e^{\pm 2i\theta}\left[  \rp-
    e^{\mp i\theta}\,A(e^{\mp i\theta}  \rx)\right]^2+
    V(e^{\mp i\theta}  \rx).
    \label{H-pm-scale}
    \ee

The boundary conditions at $x=0$ may be obtained by integrating
both sides of the eigenvalue equation for $H'$ over the interval
$[-\epsilon,\epsilon]$ and taking the limit $\epsilon\to 0$ in the
resulting expression. Doing an integration by parts, using the
fact that $A$ and $V$ are continuous functions, noting that
    \[\varphi_\epsilon(\pm\epsilon)=
    \varphi_\epsilon''(\pm\epsilon)=0,~~~~~
    \varphi_\epsilon'(\pm\epsilon)=1,~~~~~
    \gamma_\epsilon(\pm\epsilon)=\left(1\mp
    i\tan\theta\right)^{-1},\]
and introducing the notation
    \[\psi_n(0^\pm):=\lim_{x\to 0^\pm} \psi_n(x),~~~~~~
    \psi_n'(0^\pm):=\lim_{x\to 0^\pm} \psi_n'(x),\]
we find the following boundary condition at $x=0$.
    \be
    \frac{\psi_n'(0^+)}{(1-i\tan\theta)^2}-
    \frac{\psi_n'(0^-)}{(1+i\tan\theta)^2}=2i A(0)\,[
    \psi(0^+)-\psi(0^-)].
    \label{bound-zero}
    \ee
Imposing the condition that $\psi_n$ be continuous at $x=0$, i.e.,
    \be
    \psi(0^\pm)=\psi(0),
    \label{conti}
    \ee
reduces (\ref{bound-zero}) to
    \be
    e^{-2i\theta}\psi_n'(0^-)=e^{2i\theta}\psi_n'(0^+),
    \label{bound-zero-2}
    \ee
or equivalently to
    \bea
    &&|\psi_n'(0^-)|=|\psi_n'(0^+)|~~~~{\rm and}
    \label{bound-zero-2.1}\\
    &&{\rm arg}[\psi_n'(0^-)]={\rm arg}[\psi_n'(0^-)]+4\theta
    ~~~~{\rm if}~~~~\psi_n'(0^\pm)\neq 0.
    \label{bound-zero-2.2}
    \eea
Therefore, for $\psi_n$ to be differentiable at $x=0$ either
$\theta=0$ or $\psi'_n(0)=0$.

For a $PT$-invariant eigenfunction $\psi_n$, where
    \be
    \psi_n(-x)=\psi_n(x)^*,~~~~~~~~
    \psi_n'(-x)=-\psi_n'(x)^*,
    \label{PT-sym}
    \ee
and in particular
    \bea
    \psi_n(0^-)&=&\psi_n(0^+)=\psi_n(0)\in\R
    \label{real-zero}\\
    \psi'_n(0^-)&=&-\psi'_n(0^+)^*,
    \label{real-zero-der}
    \eea
(\ref{bound-zero-2.2}) implies that
    \be
    {\rm either}~~~\psi_n'(0^-)=\psi_n'(0^+)=0~~~~{\rm or}~~~~
    {\rm arg}[\psi_n'(0^\pm)]=\frac{\pi}{2}\mp 2\theta.
    \label{PT-sym-real}
    \ee
As a result $\psi_n$ is differentiable at $x=0$ if at least one of
the following conditions hold: 1.~$\psi'(0)=0$; 2.~$\theta=0$ and
$\psi'(0)$ is imaginary.\footnote{In conventional quantum
mechanics, where $\theta=0$, the $PT$-symmetric eigenfunctions of
a $PT$-symmetric Hamiltonian of the standard form $p^2+V(x)$ are
either real and even (where condition~1 holds) or imaginary and
odd (where condition~2 holds). For an example see \cite{p61}.}

Having derived the explicit expression for the boundary conditions
at $x=0$ we can identify the eigenvalue problem for the initial
Hamiltonian $H$ and the contour~(\ref{lambda-shaped}) with that of
    \be
    H'=\Lambda_-^{(0)}H_-\Lambda_-^{(0)}+
    \Lambda_+^{(0)}H_+\Lambda_+^{(0)}
    \label{H-prime-zero}
    \ee
and the requirement that the eigenfunctions belong to $L^2(\R)$
and satisfy the boundary conditions (\ref{conti}) and
(\ref{bound-zero-2}). For real eigenvalues, where we may choose to
work with the $PT$-invariant eigenfunctions, we have the boundary
conditions (\ref{real-zero}), (\ref{bound-zero-2.1}), and
(\ref{PT-sym-real}).

\section{Application to $H=p^2+x^2(ix)^{\nu}$}

For the Hamiltonians (\ref{H=}) we have
    \be
    A(x)=0,~~~~~~~~V(x)=i^\nu x^{\nu+2},~~~~~~~~
    \theta=\theta_\nu:=\frac{\pi\,\nu}{2(\nu+4)}.
    \label{H-nu}
    \ee
Inserting these in (\ref{H-pm-scale}), we are led to the following
remarkable result.
    \be
    H_\pm =e^{\pm 2i\theta_\nu}~H_{\nu+2},
    \label{H-pm-nu}
    \ee
where
    \be
    H_N:=\rp^2+|\rx|^N~~~~~{\rm for}~~~~~N\in\R.
    \label{H-N}
    \ee
Therefore, in view of (\ref{H-prime-zero}), we have
    \be
    H'=e^{-2i\theta_\nu}
    \Lambda^{(0)}_-\: H_{\nu+2}\: \Lambda^{(0)}_-+
    e^{2i\theta_\nu}
    \Lambda^{(0)}_+\: H_{\nu+2}\: \Lambda^{(0)}_+.
    \label{H-prime-nu-2}
    \ee
The eigenvalue problem for the Hamiltonian (\ref{H=}) that is
defined by the contour~(\ref{lambda-shaped}) with $\theta$ given
by (\ref{H-nu}) is equivalent to the eigenvalue equation
    \be
    e^{2i\theta_\nu{\rm sign}(\rx)}
    \left[-\psi_n''(\rx)+|\rx|^{\nu+2}\psi_n(\rx)\right]=
    E\psi_n(\rx)~~~~~{\rm for}~~~~~\rx\neq 0,
    \label{bound-eg-va}
    \ee
where $\psi_n$ are required to be continuous elements of $L^2(\R)$
satisfying
    \be
    e^{-2i\theta_\nu}
    \psi_n'(0^-)=e^{2i\theta_\nu}\psi_n'(0^+).
    \label{bound-zero-nu}
    \ee
Next, we show that the eigenfunctions $\psi_n$ never vanish at
$x=0$ and they are necessarily non-differentiable at this
point.\footnote{$\psi_n$ is necessarily twice differentiable at
all $x\neq 0$.}
    \begin{itemize}
    \item[] {\bf Lemma:} Let $\psi_n\in L^2(R)$ be a continuous
    solution of (\ref{bound-eg-va}) and (\ref{bound-zero-nu}) with
    $\nu>-2$ and $\nu\neq 0$ and $\psi_{n\pm}:\R^\pm\cup\{0\}\to\C$
    be its restrictions: $\psi_{n\pm}(\rx):=\psi_n(\rx)$ for all
    $\pm\rx\in\R^+$, $\psi_{n\pm}(0):=\psi_n(0)$, and
    $\psi'_{n\pm}(0):=\psi'_n(0^\pm)$. Then
        \be
        \psi_n(0)\neq 0\neq \psi'(0^\pm).
        \label{lemma}
        \ee
    \item[] {\bf Proof:} Clearly (\ref{bound-eg-va}) and
    (\ref{bound-zero-nu}) are respectively equivalent to
        \bea
        &&-\psi_{n\pm}''(\rx)+|\rx|^{\nu+2}\psi_{n\pm}(\rx)=
        e^{\mp2i\theta_\nu}\,
        E_n\psi_{n\pm}(\rx)~~~~{\rm for}~~~~\pm\rx\in\R^\pm,
        \label{bound-eg-va-pm}\\
        &&
        e^{-2i\theta_\nu}
        \psi_{n-}'(0)=e^{2i\theta_\nu}\psi_{n+}'(0).
        \label{bound-zero-nu-pm}
        \eea
    Multiplying both sides of (\ref{bound-eg-va-pm}) by
    $\psi_{n\pm}^*$, integrating over $\R^\pm\cup\{0\}$, and
    performing an integration by parts yield
        \be
        \pm\psi_{n\pm}(0)^*\psi'_{n\pm}(0)+
        \parallel\psi'_{n\pm}\parallel_\pm^2+
        \parallel|\rx|^{\nu/2+1}\psi_{n\pm}\parallel_\pm^2
        =e^{\mp2i\theta_\nu}E_n
        \parallel\psi_{n\pm}\parallel_\pm^2,
        \label{integral-pm}
        \ee
    where for all $\xi_\pm:\R^\pm\to\C$, $\parallel
    \xi_{\pm}\parallel_\pm^2:=\int_{\R^\pm}|\xi_\pm(\rx)|^2d\rx$.%
    \footnote{Note that because $|\rx|^{2+\nu}$ is bounded from
    below, $\parallel\psi'_{n\pm}\parallel_\pm^2$ and
    $\parallel|\rx|^{\nu/2+1}\psi_{n\pm}\parallel_\pm^2$ are
    finite numbers, \cite[\S 10.1]{Hille}.}
    Now, if at least one of $\psi(0)$, $\psi'(0^+)$, and
    $\psi'(0^-)$ vanishes, then so is the first term in
    (\ref{integral-pm}). This implies that
    $e^{\mp2i\theta_\nu}E_n$ must be real for both
    choices of the sign. For $\nu>-2$ and $\nu\neq 0$, this is only
    possible if $E_n=0$. But then the right-hand side of
    (\ref{integral-pm}) vanishes, while its left-hand side is
    strictly positive. This is a contradiction proving (\ref{lemma}).
    ~~~$\square$
    \end{itemize}
A direct implication of (\ref{lemma}) is that if $\nu>-2$ and
$\nu\neq 0$, then for all $n$, $\psi_n$ fails to be differentiable
at $\rx=0$ and that we can always normalize $\psi_n$ so that
$\psi_n(0)=1$.

For real eigenvalues $E_n$ we can take $\psi_n$ to be
$PT$-invariant and for the cases of interest, namely $\nu>0$, the
boundary conditions on the eigenvalue equation (\ref{bound-eg-va})
take the form
    \bea
    &&\psi_n(0^-)=\psi_n(0^+)\in\R,
    \label{bound-real-1}\\
    && |\psi_n'(0^-)|=|\psi_n'(0^+)|,
    \label{bound-real-2}\\
    &&{\rm arg}[\psi_n'(0^\pm)]=
    \frac{\pi}{2}\left(\frac{4+(1\mp 2)\nu}{4+\nu}\right).
    \label{bound-real-3}
    \eea

An interesting particular example is the Hamiltonian
    \be
    H=p^2-x^4,
    \label{H-pm-4}
    \ee
which corresponds to $\nu=2$ and
    \be
    H_\pm =e^{\pm \frac{i\pi}{3}}
    \left[ \rp^2+\rx^{4}\right],
    \label{H-pm-cun}
    \ee
with eigenvalue equation
    \be
    e^{\frac{i\pi}{3}\,{\rm sign}(\rx)}\left[
    -\psi_n''(\rx)+\rx^{4}\psi_n(\rx)\right]=E_n
    \psi_n(\rx)~~~~~{\rm for}~~~~~\rx\neq 0,
    \label{eg-va-even-4}
    \ee
and boundary conditions~(\ref{bound-real-1}), (\ref{bound-real-2})
and
    \be
    {\rm arg}[\psi_n'(0^\pm)]=
    \frac{(3\mp 2)\pi}{6}.
    \label{bound-real-3-4}
    \ee

The switching of the sign of the potential term from minus in
(\ref{H-pm-4}) to plus in (\ref{H-pm-cun}) and
(\ref{eg-va-even-4}) is quite remarkable. As seen from
(\ref{H-N}), (\ref{H-prime-nu-2}) and (\ref{bound-eg-va}), this is
a characteristic feature of the Hamiltonians $H$ of the
form~(\ref{H=}). In view of the discreteness of the spectrum of
the Hamiltonians $H_N$ for $N>0$, \cite{messiah}, this phenomenon
provides invaluable insight in the origin of the discreteness of
the spectrum of $H$. Indeed, as we shall show below, it leads to a
rigorous proof of the fact that for all $\nu\in (-2,\infty)$ the
spectrum of $H$ is discrete. Note that here and in what follows
the spectra of $H_N$, $H'$, and $H$ are respectively defined by
the exponentially vanishing boundary condition at $\pm\infty$
along $\R$, the latter together with the boundary conditions
(\ref{bound-real-1}) -- (\ref{bound-real-3}) at $x=0$, and
exponentially vanishing boundary condition at $\pm\infty$ along
the contour (\ref{G=}) with $\theta=\theta_\nu$.

To establish the discreteness of the spectrum of $H'$ (and
consequently $H$), we use the equivalence of the eigenvalue
problem for $H'$ with Eqs.~(\ref{bound-eg-va-pm}) and
(\ref{bound-zero-nu-pm}), and note that in terms of the functions
$y_\pm:[0,\infty)\to\C$ defined by
    \be
    y_\pm(\rx):=\psi_{n\pm}(\pm\rx),
    \label{y}
    \ee
(\ref{bound-eg-va-pm}) takes the form
    \be
    -y_\pm''(\rx)+\rx^{\nu+2}y_\pm(\rx)=\lambda_\pm y_\pm(\rx),~~~~
    {\rm for}~~~~\rx\in [0,\infty),
    \label{y=}
    \ee
where
    \be
    \lambda_{\pm}=e^{\mp 2i\theta_\nu}E_n=e^{\mp\frac{i\pi\,\nu}{\nu+4}}E_n.
    \label{E-pm}
    \ee
The eigenvalue problem for $H'$ is equivalent to finding the
solutions $y_\pm$ of (\ref{y=}) that belong to $L^2[0,\infty)$ and
satisfy
    \bea
    y_-(0)&=&y_+(0)\neq 0,
    \label{bnd1}\\
    y'_-(0)&=&-e^{4i\theta_\nu}y'_+(0)\neq 0.
    \label{bnd2}
    \eea
This problem may be treated using the classical theory of singular
boundary-value problems developed mainly by Weyl, \cite[\S
10]{Hille}. In the Appendix, we will use some basic results of
this theory to give a proof of the discreteness of spectrum of $H$
for all $\nu\in(-2,\infty)$.

We close this section by pointing out that the formulation of the
eigenvalue problem for $H$ as the differential equations
(\ref{y=}) with boundary conditions (\ref{bnd1}) and (\ref{bnd2})
is also of practical importance because it allows for the
immediate application of the known numerical, perturbative, and
variational methods that are tailored to deal with functions of a
real variable, \cite{diff-eq}. It should also be interesting to
see if one can obtain an alternative proof of the reality of the
spectrum using this formulation.

\section{Square Well Placed on a Wedge-Shaped Contour}

Consider the Hamiltonian $H=p^2+V(x)$ for the ordinary Hermitian
infinite square well potential
    \be
    V(x):=\left\{\begin{array}{ccc}
    0&{\rm for}& |x|<\frac{L}{2}\\
    \infty&{\rm for}& |x|\geq\frac{L}{2},
    \end{array}\right.
    \label{sw}
    \ee
where $L\in\R^+$. If one solves the eigenvalue problem for this
Hamiltonian on the real axis one finds an infinite discrete set of
eigenvalues
    \be
    E_n^{(0)}=\frac{\pi^2 n^2}{L^2},~~~~~~~~~n\in\Z^+.
    \label{eg-va-zero}
    \ee
As this Hamiltonian is both Hermitian and $PT$-symmetric, one may
choose to work with normalized $PT$-invariant eigenfunctions which
are given, up to an arbitrary sign, by \cite{p61}
    \be
    \psi_n^{(0)}(x)=\frac{i^{\mu_n}}{\sqrt L}\:
    \sin\left[\pi n\left(\frac{x}{L}+\frac{1}{2}\right)\right],~~~~~~~~
    \mu_n:=\frac{1+(-1)^n}{2}.
    \label{eg-fu-zero}
    \ee
We wish to explore the consequences of defining the eigenvalue
problem for the square well Hamiltonian using a wedge-shaped
contour (\ref{lambda-shaped}) with arbitrary angle
$\theta\in(0,\pi/2)$.\footnote{Taking the $\nu\to\infty$ limit of
(\ref{H=}) one obtains a similar square well Hamiltonian (with
$L=2$ and $\theta=\theta_\nu\to\pi/2$), \cite{bender-jpa-99}. For
large but finite value of $\nu$ this Hamiltonian has an infinite
number of positive real eigenvalues all of which are proportional
to $\nu^2$. Therefore in the limit $\nu\to\infty$, real part of
the spectrum is mapped to (the point at) infinity. The spectral
problem considered in this section is different from the one
treated in \cite{bender-jpa-99}, for we view the potential
(\ref{sw}) as given and take $\theta$ as a free parameter. We will
see that for large $\theta$ ($\theta>\pi/4$) the spectrum is
entirely complex.}

Pursuing the approach of Sec.~4, we find that the eigenvalue
problem for this system is equivalent to the following
boundary-value problem.
    \bea
    -\psi''_{n\pm}(\rx)&=&e^{\mp 2i\theta}
    E_n\psi_{n\pm}(\rx)~~~~
    {\rm for}~~~~
    \pm\rx\in[0,\frac{L}{2}],
    \label{eg-va-eq1}\\
    \psi_{n-}(0)&=&\psi_{n+}(0),~~~~~~~
    e^{-2i\theta}\psi'_{n-}(0)=e^{+2i\theta}\psi'_{n+}(0),
    \label{sw-bnd-zero}\\
    \psi_{n\pm}(\pm\mbox{\small$\frac{L}{2}$})&=&0.
    \label{sw-bnd}
    \eea
Clearly $\psi_{n\pm}$ determine the eigenfunctions $\psi_n$ of the
system according to
    \be
    \psi_n(\rx)=\psi_{n\pm}(\rx),~~~{\rm if}~~~
    \pm \rx\in[0,\mbox{\small$\frac{L}{2}$}].
    \label{psi-n=}
    \ee
They belong to
    \be
    {\cal H}':=\left\{\psi\in\left.
    L^2[-\mbox{\small$\frac{L}{2}$},\mbox{\small$\frac{L}{2}$}]\,
    \right|\,\psi(\pm \mbox{\small$\frac{L}{2}$})=0\right\}.
    \label{Hilbert-prime}
    \ee

The eigenvalue problem~(\ref{eg-va-eq1}) -- (\ref{sw-bnd}) may be
easily solved: Zero is an acceptable eigenvalue only for
$\theta=\pi/4$. The corresponding $PT$-invariant eigenfunction is
given by
    \be
    \psi(\rx)=\pm c\left(\rx\mp\mbox{\small
    $\frac{L}{2}$}\right)~~~~
    {\rm for}~~~~\pm\rx\in[0,\mbox{\small $\frac{L}{2}$}],
    \label{E=0}
    \ee
where $c$ is a real normalization constant. The eigenfunctions
with nonzero eigenvalues have the form
    \be
    \psi_{n\pm}(\rx)=c_\pm\,e^{i\omega_{n\pm}\rx}+d_\pm\,
    e^{-i\omega_{n\pm}\rx},
    \label{sw=psi-n=}
    \ee
where $\omega_{n\pm}:=e^{\mp i\theta}\sqrt{E_n}$ and
    \bea
    &&c_-=\frac{1}{2}\left[(1+e^{2i\theta})c_+
    +(1-e^{2i\theta})d_+\right],~~~~~~
    d_-=\frac{1}{2}\left[(1-e^{2i\theta})c_+
    +(1+e^{2i\theta})d_+\right],
    \label{cd-m}\\
    &&c_+ e^{i\omega_{n+}L/2}+d_+ e^{-i\omega_{n+}L/2}=0,~~~~~~~
    c_-e^{-i\omega_{n+}L/2}+d_-e^{i\omega_{n+}L/2}=0.
    \label{cd-p}
    \eea
Eqs.~(\ref{cd-m}) and (\ref{cd-p}) follow from the boundary
conditions (\ref{sw-bnd-zero}) and (\ref{sw-bnd}), respectively.
They have a nontrivial solution provided that the eigenvalues
$E_n$ satisfy a transcendental equation that takes the following
simple form in terms of the variable
$u_n:=\cos(\theta)L\sqrt{E_n}$,
    \be
    \tan(\theta)\sinh[\tan(\theta)u_n]=\sin(u_n).
    \label{secular}
    \ee
For $\theta=0$ it reduces to $\sin(u_n)=0$, and one recovers
$E_n=E_n^{(0)}$. But for $\theta>0$ it has a finite number
$N(\theta)$ of real solutions where $N$ is a decreasing function
of $\theta$. In particular, for $\theta>\pi/4$, $N(\theta)=0$ and
there is no real solution. As one decreases the value of $\theta$
from $\pi/2$ down to zero one encounters an infinite strictly
increasing sequence $\{ {\cal E}_\ell\}$ of exceptional points
\cite{exceptional}. The angles $\theta$ for the corresponding
wedge-shaped contours form a strictly decreasing sequence
$\{\theta_\ell\}$ that converges to zero. Table~\ref{tab1} lists
the values of the first five exceptional points and the
corresponding angles $\theta_\ell$.
    \begin{table}[ht]
    \vspace{.5cm}
  \begin{center}
  \begin{tabular}{||c||c|c|c|c|c||}
  \hline \hline
  $\ell$ & 1 & 2 & 3 & 4 & 5 \\
  \hline\hline
  ${\cal E}_\ell$ & 0 & $61.58~L^{-2}$& 200.9~$L^{-2}$ &
  418.9~$L^{-2}$ & 715.7~$L^{-2}$\\
  \hline
  $\theta_\ell$ & 45.00$^\circ$ & 14.81$^\circ$ & 9.88$^\circ$
  & 7.59$^\circ$ & 6.23$^\circ$ \\
  \hline \hline\end{tabular}
  \end{center}
    \centerline{\parbox{10cm}{\caption{The first five exceptional points
    ${\cal E}_\ell$ and the corresponding exceptional values
    $\theta_\ell$ of $\theta$.}
    \label{tab1}}}
    \end{table}

In general, the number of real eigenvalues are given by
    \be
    N(\theta)=\left\{\begin{array}{ccc}
     2\ell-1 & {\rm for}& \theta\in(\theta_{\ell+1},\theta_{\ell})
        ~{\rm with}~\ell\geq 1\\
     2\ell-2 & {\rm for}& \theta=\theta_\ell~{\rm with}~\ell\geq 2\\
     1& {\rm for}& \theta=\theta_1=\pi/4.
    \end{array}\right.
    \label{N=}
    \ee
Because the eigenvalues are nondegenerate, the dimension of the
invariant subspace spanned by the eigenfunctions with a real
eigenvalue is $N(\theta)$. This $N(\theta)$-dimensional subspace
is the underlying vector space ${\cal V}$ for both the reference
Hilbert space (${\cal H}$) and the physical Hilbert space (${\cal
H}_{\rm phys}$) of the system. For $\theta=0$, $N(\theta)=\infty$
and ${\cal H}$, ${\cal H}_{\rm phys}$, and ${\cal H}'$ coincide.
But for $\theta>0$, ${\cal V}$ is finite-dimensional. In
particular, for $\theta>\theta_1=\pi/4$ the vector space ${\cal
V}$ is zero-dimensional and the system does not admit a unitary
quantum description.

Another peculiar feature of this system is that the dimension of
the physical Hilbert space takes even values only for the
exceptional values $\theta_\ell$ of $\theta$ with $\ell\geq 2$. As
these constitute a measure zero subset of $[0,\pi/4)$, the
physical Hilbert space is generically odd-dimensional!

Three comments are in order.
\begin{enumerate}
    \item If one defines the eigenvalue problem using the Neumann
boundary conditions at $\rx=\pm L/2$, i.e., requires
$\psi_{n\pm}'(\pm L/2)=0$, the (nonzero) eigenvalues are given by
Eq.~(\ref{secular}) with the sign of the right-hand side changed.
The corresponding pseudo-Hermitian quantum system shares the
general features of the square well system discussed above. The
only difference is that for all values of $\theta$, zero is an
eigenvalue with a constant eigenfunction. In particular the
physical Hilbert space is finite-dimensional for
$0<\theta\leq\pi/2$, infinite-dimensional for $\theta=0$, and
one-dimensional for $\pi/4\leq\theta<\pi/2$.
    \item The quantum system corresponding to the square well
Hamiltonian placed on a wedge-shaped contour defines a
$PT$-symmetric quantum system which is fundamentally different
from the $PT$-symmetric square well studied in
\cite{sw-znojil,bagchi-quesne,p61}. The latter system involves a
non-Hermiticity parameter $Z\in[0,\infty)$. As one increases the
value of $Z$ (starting from zero) one encounters an infinite
sequence of exceptional points which correspond to a strictly
increasing sequence $\{Z_\ell\}$ of exceptional values of $Z$. As
a result unlike the system introduced above the physical Hilbert
space is always infinite-dimensional. In particular, for $0\leq
Z<Z_1$ the reference Hilbert space ${\cal H}$ coincides with
${\cal H}'$.
    \item For the square well system defined on a wedge-shaped
contour, $\theta=0$ ----- which corresponds to the Hermitian limit
of the problem ----- is an accumulation point of the exceptional
values $\theta_\ell$ of $\theta$. This is the reason why for all
positive values of $\theta$ the physical Hilbert space is
finite-dimensional.\footnote{It is not difficult to see that the
same holds for negative $\theta$.} This observation shows that
changing the domain of the definition of a Hamiltonian from the
real line to a complex contour can lead to completely different
quantum systems. For example, for $\theta=\theta_2$ the physical
Hilbert space is two-dimensional. Therefore it describes the
interaction of a spin-half particle with a magnetic field
\cite{critique}. In contrast, for $\theta=0$, the system describes
the one-dimensional motion of a particle that is trapped between
two impenetrable walls.
    \end{enumerate}

\section{Application of Pseudo-Hermitian QM for Square Well along
the Wedge-Shaped Contour with $\theta=\theta_2$}

The largest value of the angle $\theta$ that corresponds to a
nontrivial unitary quantum system is $\theta=\theta_2\approx
14.81^\circ$. For this choice of $\theta$ the Hamiltonian has two
real eigenvalues. They are $E_1\approx 9.09 L^{-2}$ and $E_2={\cal
E}_2\approx 61.6 L^{-2}$. The corresponding eigenfunctions
$\psi_1$ and $\psi_2$ are given by (\ref{psi-n=}) and
(\ref{sw=psi-n=}) where
    \bea
    && c_{n\pm}=c_{\pm}(E_n),~~~~~~~~~~~~~~~d_{n\pm}=d_\pm(E_n),
    \label{e1}\\
    && c_\pm(E):=\frac{\cun(E)}{1-e^{\pm i\Omega_\pm(E)}},
    ~~~~~~~~
       d_\pm(E):=\frac{\cun(E)}{1-e^{\mp i\Omega_\pm(E)}},
    \label{e2}\\
    &&\Omega_\pm(E):=e^{\mp i \theta}L\sqrt E,
    \label{e3}
    \eea
and $\cun(E)$ is an arbitrary real normalization constant.
Substituting (\ref{e1}) and (\ref{e2}) in (\ref{sw=psi-n=}) and
using (\ref{psi-n=}), we have
    \be
    \psi_n(\rx)=\psi_{n\pm}(\rx)=\frac{\cun_n~\sin\left[\Omega_\pm(E_n)
    \left(\frac{1}{2}\mp\frac{\rx}{L}\right)\right]}{
    \sin\left[\frac{\Omega_\pm(E_n)}{2}\right]}~~~~~~~
    {\rm for}~~~~~~\pm\rx\in[0,\mbox{\small$\frac{L}{2}$}],
    \label{psi=sw=}
    \ee
where $\cun_n\in\R^+$ are normalization constants, and $n=1,2$.

The underlying vector space ${\cal V}$ for the reference and the
physical Hilbert spaces is the two-dimensional subspace of ${\cal
H}'$ spanned by $\psi_1$ and $\psi_2$. The reference Hilbert space
${\cal H}$ is obtained by endowing ${\cal V}$ with the subspace
inner product $\br\cdot|\cdot\kt$ induced from ${\cal H}'$.
Choosing the normalization constants as $\cun_1\approx 1.226
L^{-1/2}\kappa$ and $\cun_2\approx 0.717 L^{-1/2}\kappa$, for some
$\kappa\in\R^+$, we have
    \be
    \br\psi_1|\psi_1\kt=\br\psi_2|\psi_2\kt=\kappa^2,~~~~~~
    \br\psi_1|\psi_2\kt=\br\psi_2|\psi_1\kt=r\kappa^2,
    \label{normalization}
    \ee
where
    \be
        r\approx 0.068.
    \ee
Clearly $\{\psi_1,\psi_2\}$ form a non-orthogonal basis of ${\cal
H}$. We can use the Gram-Schmidt procedure \cite{linear-algebra}
to construct an orthonormal basis
$\{\varepsilon_1,\varepsilon_2\}$ according to
    \be
    \varepsilon_1:=\kappa^{-1}\psi_1,~~~~~~~~~
    \varepsilon_2:=\frac{\psi_2-r\psi_1}{\kappa\sqrt{1-r^2}}.
    \label{basis}
    \ee
In this basis the Hamiltonian is represented by the following
manifestly non-Hermitian $2\times 2$ matrix
    \be
    \tilde H=\left( \begin{array}{cc}
    E_1 & \frac{r(E_2-E_1)}{\sqrt{1-r^2}}\\
    0 & E_2\end{array}\right)\approx L^{-2}
    \left(\begin{array}{cc}
    9.09 & 3.56\\
    0 & 61.6\end{array}\right).
    \label{tilde-H}
    \ee

We can compute the adjoint of $H$ using its matrix
representation~(\ref{tilde-H}) and determine its eigenvectors
$\phi_n$ that together with $\psi_n$ form a biorthonormal system
for ${\cal H}$. This yields
    \be
    \phi_1=\kappa^{-1}\left(\varepsilon_1-\frac{r}{\sqrt{1-r^2}}
    \,\varepsilon_2\right)=
    \frac{\psi_1-r\psi_2}{\kappa^2(1-r^2)},~~~~~~
    \phi_2=\frac{\varepsilon_2}{\kappa\sqrt{1-r^2}}=
    \frac{\psi_2-r\psi_1}{\kappa^2(1-r^2)}.
    \label{phi}
    \ee
Now, we are in a position to compute the metric operator $\eta_+$.
In view of (\ref{eta+}) and (\ref{phi}), it has the following
matrix representation in the orthonormal basis $\{\varepsilon_1,
\varepsilon_2\}$.
    \be
    \tilde\eta_+=\kappa^{-2}\left( \begin{array}{cc}
    1 & -\frac{r}{\sqrt{1-r^2}}\\
    -\frac{r}{\sqrt{1-r^2}} & \frac{1+r^2}{1-r^2}
    \end{array}\right)\approx\kappa^{-2}\left(\begin{array}{cc}
    1 & -0.068\\
    -0.068 & 1.009\end{array}\right).
    \label{tilde-eta+}
    \ee
In view of this relation, we have, for all $\xi,\zeta\in{\cal V}$,
    \bea
    \br\xi,\zeta\kt_+:=\br\xi|\eta_+\zeta\kt&=&\kappa^{-2}\left[
    \xi_1^*\zeta_1-\frac{r
    \left(\xi_1^*\zeta_2+\xi_2^*\zeta_1\right)}{\sqrt{1-r^2}}+
    \frac{(1+r^2)\xi_2^*\zeta_2}{1-r^2}\right]\nn\\
    &\approx&\kappa^{-2}\left[\xi_1^*\zeta_1-0.068\left(
    \xi_1^*\zeta_2+\xi_2^*\zeta_1\right)
    +1.009\,\xi_2^*\zeta_2\right],
    \label{inner-prod-sw}
    \eea
where $\xi_n=\br\varepsilon_n|\xi\kt$,
$\zeta_n=\br\varepsilon_n|\zeta\kt$, and $n=1,2$. Note that the
coefficient $\kappa^{-2}$ is a trivial scaling of the inner
product.

If we use (\ref{inner-prod-sw}) to compute the inner product of
the eigenvectors $\psi_n$, we find that as expected
$\{\psi_1,\psi_2\}$ form an orthonormal basis of the physical
Hilbert space, $\br\psi_n,\psi_m\kt_+=\delta_{mn}$ for $m,n=1,2$.
This also shows that the Hamiltonian viewed as acting in ${\cal
H}_{\rm phys}$ is a Hermitian operator.

Next, we construct the physical observables $O$ of the system.
This requires the computation of $\rho=\sqrt\eta_+$. The matrix
representation of $\rho$ in the basis $\{\varepsilon_1,
\varepsilon_2\}$ has the form
    \be
    \tilde\rho=\sqrt{\tilde\eta_+}\approx\kappa^{-1}
    \left(\begin{array}{cc}
    0.999&-0.034\\
    -0.034&1.004\end{array}\right).
    \label{rho-sw=}
    \ee
According to (\ref{observable}), the physical observables are
given by $O=\sum_{\ell=0}^3 \omega_\ell \Sigma_\ell$ where
$\omega_\ell\in\R$ are arbitrary constants, $\Sigma_0$ is the
identity operator acting in ${\cal H}$, for $\ell=1,2,3$,
$\Sigma_\ell$ are defined through their matrix representations in
the basis $\{\varepsilon_1, \varepsilon_2\}$ according to
    \be
    \tilde\Sigma_\ell=\tilde\rho^{-1}\sigma_\ell\tilde\rho,
    \label{matrix}
    \ee
and $\sigma_\ell$ are Pauli matrices. Specifically,
    \bea
    \tilde\Sigma_0&=&\sigma_0:=\left(\begin{array}{cc}
    1 & 0\\
    0 & 1\end{array}\right),~~~~
    \tilde\Sigma_1\approx\left(\begin{array}{cc}
    0 & 1.005\\
    0.995 & 0\end{array}\right),\nn\\
    \tilde\Sigma_2&\approx&i\left(\begin{array}{cc}
    0.068 & -1.007\\
    0.998 & -0.068\end{array}\right),~~~~
    \tilde\Sigma_3\approx\left(\begin{array}{cc}
    1.002 &-0.068\\
    0.068 &-1.002\end{array}\right).\nn
    \eea
Using these relations and (\ref{tilde-H}), we can show that indeed
    \be
    H\approx L^{-2}\left(35.5\: \Sigma_0+
    1.78\:\Sigma_1-26.2\:\Sigma_3\right).
    \label{H-tilde=}
    \ee

Next, we compute the Hermitian Hamiltonian $h$ of
(\ref{hermitian-h}) that is associated with $H$. We can obtain the
matrix representation $\tilde h$ of $h$ in the basis
$\{\varepsilon_1, \varepsilon_2\}$ using either of (\ref{rho-sw=})
and (\ref{tilde-H}) or (\ref{matrix}) and (\ref{hermitian-h}).
Both yield
    \be
    \tilde h\approx L^{-2}
    \left(\begin{array}{cc}
    9.15 &1.78\\
    1.78 &61.5\end{array}\right)=L^{-2}(
    35.5\:\sigma_0+1.78\:\sigma_1-26.2\:\sigma_3).
    \label{tilde-h-Hermitian}
    \ee
Therefore,
    \[h\approx L^{-2}\left(9.15\,|\varepsilon_1\kt\br\varepsilon_1|+
    1.78\,(|\varepsilon_1\kt\br\varepsilon_2|+
    |\varepsilon_2\kt\br\varepsilon_1|)+61.5\,
    |\varepsilon_2\kt\br\varepsilon_2|\right).\]

Having obtained the biorthonormal system
$\{|\psi_n\kt,|\phi_n\kt\}$, we can also compute the generalized
parity ${\cal P}$, time-reversal ${\cal T}$, and
charge-conjugation ${\cal C}$ operators of \cite{jmp-2003},
namely\footnote{See also \cite{ahmed}.}
    \bea
    {\cal P}&:=&|\phi_1\kt\br\phi_1|-|\phi_2\kt\br\phi_2|,
    \label{P=}\\
    {\cal T}&:=&|\phi_1\kt\star\br\phi_1|-|\phi_2\kt\star\br\phi_2|,
    \label{t=}\\
    {\cal C}&:=&|\psi_1\kt\br\phi_1|-|\psi_2\kt\br\phi_2|,
    \eea
where $\star$ is the complex-conjugation defined by
    \be
    \star\,|\zeta\kt:=\sum_{n=1}^2 \br\varepsilon_n|\zeta\kt^*
    |\varepsilon_n\kt=\sum_{n=1}^2 \br\zeta|\varepsilon_n\kt\:
    |\varepsilon_n\kt,
    ~~~~~~{\rm for~all}~~~~~\zeta\in{\cal H}.
    \label{star}
    \ee
In particular, in the basis $\{\varepsilon_1, \varepsilon_2\}$,
$\star$ is represented by ordinary complex conjugation `$*$' of
complex vectors,
    \be
    *\,\vec z:={\vec z}^*,~~~~{\rm where}~~~~
    \vec z=\left(\begin{array}{c}
    \br\varepsilon_1|\zeta\kt\\
    \br\varepsilon_2|\zeta\kt\end{array}\right)\in\C^2,~~~~~
    \zeta\in{\cal H}.
    \ee

As explained in \cite{jmp-2003}, unlike ${\cal C}$ which is always
an involution (${\cal C}^2=1$), ${\cal P}$ and ${\cal T}$ need not
be involutions. Requiring them to be involutions restricts the
choice of the biorthonormal system. In the case at hand, this
restriction amounts to fixing the normalization constant for the
eigenvectors $\psi_n$ as
    \be
        \kappa=(1-r^2)^{-1/4}\approx 1.001.
    \ee
Making this choice, we find that the matrix representations of
${\cal P}$, ${\cal T}$, and ${\cal C}$, in the basis
$\{\varepsilon_1, \varepsilon_2\}$, are respectively given by
    \bea
    \tilde{\cal P}&=&\left(\begin{array}{cc}
    \sqrt{1-r^2} & -r\\
    -r & -\sqrt{1-r^2}\end{array}\right)\approx
    \left(\begin{array}{cc}
    0.998 &-0.068\\
    -0.068 & -0.998\end{array}\right),
    \label{tilde-P=}\\
    \tilde{\cal T}&=&\tilde{\cal P}*,~~~~~~~~~~~~
    \tilde{\cal C}=\left(\begin{array}{cc}
    1 & -\frac{2r}{\sqrt{1-r^2}}\\
    0 & -1\end{array}\right)\approx
    \left(\begin{array}{cc}
    1 &-0.136\\
    0 & -1\end{array}\right).
    \label{tilde-C=}
    \eea
Using these relations we can directly check that indeed
    \be
    {\cal P}^2={\cal T}^2={\cal C}^2=1,
    ~~~~~~{\cal C}=\eta_+^{-1}{\cal P},~~~~~~
    [H,{\cal C}]=[H,{\cal PT}]=0.
    \label{sym}
    \ee
In view of the identity ${\cal PT}=\star$, the ${\cal
PT}$-symmetry of $H$ corresponds to the fact that $H$ is a real
operator with respect to the complex-conjugation (\ref{star}),
i.e., $\star\,H\,\star=H$. An explicit manifestation of the latter
relation is that $\tilde H$ is a real matrix.\footnote{Because the
matrix representation $\tilde H$ of the Hamiltonian is not
symmetric, the definition of observables proposed in
\cite{bbj-erratum} cannot be employed, \cite{comment}.}

\section{Formulation Based on the ${\cal CPT}$-Inner Product,
Discussion, and Conclusion}

In this paper we have presented a formulation of $PT$-symmetric
theories defined along a complex contour in which the state
vectors belong to the familiar Hilbert space of square-integrable
functions. This formulation has a number of advantages. Firstly,
it yields the necessary means for a straightforward application of
the results of the theory of pseudo-Hermitian operators. Secondly,
it provides a novel description of the Hamiltonians of the form
(\ref{H=}) that reveals the origin of the discreteness of their
spectrum. Finally, it is practically appealing for it allows for a
direct application of the standard approximation schemes developed
for solving differential equations on the real line
\cite{diff-eq}.

In order to elucidate the practical aspects of our method we have
considered the $PT$-symmetric system obtained by placing an
infinite square well potential on a wedge-shaped contour $\Gamma$.
We have conducted a comprehensive study of this model showing that
as soon as one makes the characteristic angle $\theta$ of the
contour $\Gamma$ different from zero (i.e., moves off the real
axis) the physical Hilbert space of the system becomes
finite-dimensional. The dimension of this space depends on
$\theta$. It changes at certain critical values of $\theta$ that
correspond to the exceptional spectral points associated with the
system. The simplest nontrivial case occurs at the second
exceptional point where $\theta\approx 14.81^\circ$ and the
physical Hilbert space is two-dimensional. For this case we showed
how one could employ the constructions developed in the framework
of pseudo-Hermitian quantum mechanics to determine the explicit
form of the inner product of the physical Hilbert space, the
physical observables, and the corresponding Hermitian Hamiltonian.

The results reported in this paper show that $PT$-symmetric
quantum mechanics is indeed a special case of pseudo-Hermitian
quantum mechanics. In order to apply the pseudo-Hermitian quantum
mechanics to $PT$-symmetric systems defined on a complex contour,
one may employ the fact that these systems admit a convenient
description in terms of $PT$-symmetric Hamiltonians defined on the
real line. The latter may be treated most perspicuously within the
framework of pseudo-Hermitian quantum mechanics. In particular,
one can compute the observables of the theory and explore its
classical limit as outlined in \cite{p61,p64}.

There is also a more direct, but less practical, pseudo-Hermitian
description of $PT$-symmetric systems defined on a complex contour
$\Gamma$. This is also suggested by the analysis of
Sec.~2.\footnote{See also \cite{dorey}.} It involves identifying
the reference Hilbert space ${\cal H}$ with $L^2(\Gamma)$, where
the contour $\Gamma$ is viewed as a one-dimensional real
submanifold of $\R^2=\C$, i.e., a continuous (piecewise regular)
plane curve. The relationship between this `complex
pseudo-Hermitian description' and the `real pseudo-Hermitian
description' that is based on transforming the system onto the
real line may be reduced to the action of a diffeomorphism ${\cal
G}$ of the complex plane that maps the real axis onto the contour
$\Gamma$. This mapping may be identified with the arc-length
parametrization of $\Gamma$. In view of (\ref{zeta}), we can
parameterize $\Gamma$ by the $x$-coordinate. We can use this
parametrization to define the arc-length parameter: $\rx={\cal
F}(x):=\int_0^x \sqrt{1+f'(s)^2}\,ds$. Note that for the contours
of interest ${\cal F}:\R\to\R$ is a diffeomorphism. The
restriction of ${\cal G}$ onto the real axis defines the following
mapping of $\R$ onto $\Gamma$.
    \be
    {\cal G}(\rx):=x+i f(x)=
    {\cal F}^{-1}(\rx)+if({\cal F}^{-1}(\rx)),~~~~~~{\rm for~all}
    ~~~~~~~\rx\in\R.
    \label{diffeo=}
    \ee
This in turn induces a unitary operator $u_{_{\cal G}}:L^2(\R)\to
L^2(\Gamma)$ defined by\footnote{One might try to express
$u_{_{\cal G}}$ in the form $e^{i\{G(\rx),\rp\}/2}$ for some
complex-valued function $G$ by extending the results of
\cite{diffeo}.}
    \be
    (u_{_{\cal G}}\psi)(z):=\psi({\cal G}^{-1}(z)),~~~~~~~
    {\rm for~all}~~~\psi\in L^2(\R),~~~z\in\Gamma.
    \label{u-G}
    \ee
Alternatively, setting $\Psi:=u_{_{\cal G}}\psi$ we have
    \be
    \Psi(z)=\psi(\rx)~~~~~~~~{\rm if~and~only~if}~~~~~~~~~
    z={\cal G}(\rx).
    \label{psi=psi}
    \ee

The statement that $u_{_{\cal G}}$ is a unitary operator means
that for all $\psi,\phi\in L^2(\R)$
    \be
    \br u_{_{\cal G}}\psi|u_{_{\cal G}}\phi\kt_\Gamma=
    \br\psi|\phi\kt,
    \label{unitary=}
    \ee
where $\br\cdot|\cdot\kt_\Gamma$ is the inner product of
$L^2(\Gamma)$, i.e.,
    \be
    \br\Psi|\Phi\kt_\Gamma:=\int_\Gamma \Psi(z)^*\Phi(z)\,dz.
    \label{L-2-Gamma}
    \ee
The validity of Eq.~(\ref{unitary=}) becomes obvious once we
identify $\Gamma$ with a plane curve and view the right-hand side
of (\ref{L-2-Gamma}) as a line integral. Letting $\Psi:= u_{_{\cal
G}}\psi$ and $\Phi:= u_{_{\cal G}}\phi$ and using (\ref{diffeo=})
and (\ref{u-G}), we have
    \[\br u_{_{\cal G}}\psi|u_{_{\cal G}}\phi\kt_\Gamma=
    \br\Psi|\Phi\kt_\Gamma=\int_{\R}
    \Psi({\cal G}(\rx))^*\Phi({\cal G}(\rx))\,d\rx=
    \int_{\R} \psi(\rx)^*\phi(\rx)\,d\rx=\br\psi|\phi\kt.\]

An important property of $u_{_{\cal G}}$ is that it establishes a
one-to-one correspondence between the ingredients of the two
pseudo-Hermitian descriptions of the system; to each linear
operator $A$ acting in $L^2(\R)$ it associated a linear operator
${\cal A}:=u_{_{\cal G}} A u_{_{\cal G}}^{-1}$ acting in
$L^2(\Gamma)$. In particular, it maps the charge-conjugation
operator $C:=\eta^{-1}P$ of the real description to the
charge-conjugation operator ${\cal C}:L^2(\Gamma)\to L^2(\Gamma)$
of the complex description according to
    \be
    {\cal C}:=u_{_{\cal G}} C u_{_{\cal G}}^{-1}.
    \label{cc}
    \ee
In view of the results of \cite{jmp-2003}, for the Hamiltonians
(\ref{H=}) with $\nu\geq 0$, the operator ${\cal C}$ is nothing
but the charge-conjugation operator introduced in \cite{bbj}. In
fact, what the authors of \cite{bbj} do is to define $C$ on the
real line (though they use the same symbol for both $C$ and ${\cal
C}$), perform the diffeomorphism $u_{_{\cal G}}$ to obtain ${\cal
C}$, and then use it in a contour integral along $\Gamma$ to
define their ${\cal CPT}$-inner product:
    \be
    \br\Psi,\Phi\kt_{\cal CPT}:=\int_\Gamma
    [{\cal CPT}\Psi(z)]\Phi(z)\,dz~~~~~{\rm for}~~~~~
    \Psi,\Phi\in L^2(\Gamma).
    \label{cpt-complex}
    \ee
Note that in the real description \cite{jmp-2003},
    \be
    \br \psi,\phi\kt_{CPT}:=\int_\R
    [{CPT}\psi(\rx)]\phi(\rx)d\rx=
    \br\psi|\eta_+|\phi\kt=\br\psi,\phi\kt_+~~~~~{\rm for}~~~~~
    \psi,\phi\in L^2(\R).
    \label{cpt-real}
    \ee
Moreover, the eigenfunctions $\Psi_n$ (respectively
$\psi_n$)\footnote{Recall that according to the analysis of
Sec.~5, the eigenfunctions $\psi_n$ and $\Psi_n$ are related via
$\Psi_n({\cal G}(\rx))=\psi_n(\rx)$.} form an orthonormal set with
respect to $\br\cdot,\cdot\kt_{\cal CPT}$ (respectively
$\br\cdot,\cdot\kt_+$),
    \be
    \br\Psi_m,\Psi_n\kt_{\cal CPT}=\delta_{mn}=
    \br\psi_m,\psi_n\kt_+=\br\psi_m|\eta_+|\psi_n\kt.
    \label{inn-complex}
    \ee

Next, we introduce a metric operator $\eta_+^{\C}:L^2(\Gamma)\to
L^2(\Gamma)$ and the corresponding inner product
$\br\cdot,\cdot\kt_+^{\C}:L^2(\Gamma)\times L^2(\Gamma)\to\C$
according to
    \be
    \eta_+^{\C}:=u_{_{\cal G}}\eta_+u_{_{\cal G}}^{-1},
    ~~~~~~~~~
    \br\cdot,\cdot\kt_+^{\C}:=\br\cdot|\eta_+^{\C}\cdot\kt_\Gamma.
    \label{complex-eta}
    \ee
In view of the identity $\Psi_n=u_{_{\cal G}}\psi_n$ and the fact
that $u_{_{\cal G}}$ is unitary, we then find
    \be
    \delta_{mn}=\br\psi_m|\eta_+|\psi_n\kt=\br u_{_{\cal G}}^{-1}
    \Psi_m|\eta_+|u_{_{\cal G}}^{-1}\psi_n\kt=
    \br\Psi_m|\eta^{\C}_+|\Psi_n\kt.
    \label{inn-complex-2}
    \ee
Equations (\ref{inn-complex}) and (\ref{inn-complex-2}) show that
$\Psi_n$, which are supposed to form a complete set, are
orthonormal with respect to both the ${\cal CPT}$-inner product
(\ref{cpt-complex}) and the inner product
$\br\cdot,\cdot\kt_+^{\C}$. This proves that these two inner
products are identical. Therefore, the formulation of
$PT$-symmetric quantum mechanics based on the ${\cal CPT}$-inner
product, as outlined in \cite{bbj}, admits a complete description
in terms of the theory of pseudo-Hermitian operators.

\vspace{.8cm}


\noindent{\large\textbf{Note:}} After the completion of this
project, I discovered a preprint of Znojil~\cite{znojil-2004}
where he considers the analytic continuation of the $PT$-symmetric
square well of Ref.~\cite{sw-znojil} onto a smooth complex
contour. The spectral properties of this system is similar to the
one considered in Sec.~6. In both cases the spectrum is determined
through a set of boundary conditions at the intersection point of
the contour and the imaginary axis. The main difference between
the two systems is that the defining boundary conditions used in
\cite{znojil-2004} are postulated whereas those used in Sec.~6 are
derived. As explained in Sec.~4, the latter are the general
boundary conditions associated with the wedge-shaped contours.

\section*{Acknowledgment}

I would like to thank Varga Kalantarov for useful discussions and
Carl Bender whose criticism of my earlier work motivated the
present study.

\section*{Appendix: Discreteness of the Spectrum of
(\ref{H=})}

 \begin{itemize}
    \item[] {\bf Theorem:} The spectrum of the Hamiltonians
$H=p^2+x^2(ix)^\nu$ defined by the contour (\ref{lambda-shaped})
with $\theta=\theta_\nu:=\pi\nu/[2(\nu+4)]$ is discrete for all
$\nu\in(-2,\infty)$.
    \item[] {\bf Proof:}
For $\nu=0$ this statement is well-known to hold \cite{messiah}.
To prove it for $\nu\neq 0$, we prove the equivalent statement
that for all $\nu\in(-2,\infty)$ the following boundary-value
problem has a solution only for a discrete set of values of
$E_n$.\footnote{The equivalence of this statement with that of the
above theorem is established in Sec.~5. $E_n$ are the eigenvalues
of $H$.}
    \bea
    &&-y_\pm''(\rx)+\rx^{\nu+2}y_\pm(\rx)=\lambda_\pm y_\pm(\rx)
    ~~~~{\rm for}~~~~\rx\in[0,\infty),
    \label{z-y=}\\
    &&\lambda_{\pm}=e^{\mp 2i\theta_\nu}E_n\in\C,
    \label{z-E-pm}\\
    &&y_\pm\in L^2[0,\infty),
    \label{z-bnd0}\\
    &&y_-(0)=y_+(0)\neq 0,
    \label{z-bnd1}\\
    &&y'_-(0)=-e^{4i\theta_\nu}y'_+(0)\neq 0.
    \label{z-bnd2}
    \eea

Let $\lambda\in\C$ be arbitrary, and consider finding solutions
$y(\cdot;\lambda)$ of
    \be
    -y''(\rx)+\rx^{\nu+2}y(\rx)=
    \lambda\, y(\rx),~~~~
    {\rm with}~~~~\nu>-2,~~\rx\in[0,\infty),
    \label{w1}
    \ee
that belong to $L^2[0,\infty)$. Then because $\rx^{\nu+2}$ is
bounded below by zero, one has the so-called limit point case
\cite[\S 10.1]{Hille} where there is at most one linearly
independent $L^2$-solution and such a solution exists for all
non-real $\lambda$ and has the form
    \be
    y(\rx;\lambda)=C(\lambda)[y_1(\rx;\lambda)+m(\lambda)
    y_2(\rx;\lambda)],
    \label{w2}
    \ee
where $C(\lambda)\in\C-\{0\}$ is a constant, $y_1$ and $y_2$ are
the fundamental solutions of (\ref{w1}) satisfying
    \be
    y_1(0;\lambda)=0,~~~y_1'(0;\lambda)=-1,~~~
    y_2(0;\lambda)=1,~~~y_1'(0;\lambda)=0,
    \label{bound-1-2}
    \ee
and $m:\C\to\C$ is a function having the property \cite[\S
10.2]{Hille}
    \be
    m(\lambda^*)=m(\lambda)^*.
    \label{m=}
    \ee

Now, consider the boundary-value problem: (\ref{w1}), $y'(0)=0$,
and $y\in L^2[0,\infty)$. Because $\rx^{\nu+2}\to\infty$ as
$\rx\to\infty$, this problem defines a discrete (pure point)
spectrum ${\cal S}:=\{\lambda_k|k\in\Z^+\}$ which is real and
unbounded, \cite[\S 10.3]{Hille}. Furthermore, the eigenfunction
associated with $\lambda_k$ is, up to a multiplicative constant,
$y_2(\cdot\,;\lambda_k)$, and the function $m$ has the following
spectral resolution:
    \be
    m(\lambda)=\sum_{k=1}^\infty
    \frac{\sigma_k}{\lambda_k-\lambda},
    \label{mm=}
    \ee
where $\sigma_k=\left[\int_0^\infty|y_2(\rx;\lambda_k)|^2
d\rx\right]^{-1}\in\R$. In particular, $m$ is a holomorphic
function in $\C-{\cal S}$ and $\lambda_k$ are the poles of $m$
which are all simple.\footnote{Note that $\lambda_k>0$ for all
$k\in\Z^+$ and that ${\cal S}$ has no accumulation (cluster)
point.}

Next, consider the following two possibilities:
    \begin{enumerate}
\item $\lambda_+\in\R$ or $\lambda_-\in\R$: First suppose
$\lambda_+\in\R$, then $\lambda_-\notin\R$ and we have
    \be
    y_-(\rx;\lambda_-)=C(\lambda_-)[y_1(\rx;\lambda_-)+m(\lambda_-)
    y_2(\rx;\lambda_-)],
    \label{w2-minus}
    \ee
where $m$ is given by (\ref{mm=}). Eqs.~(\ref{z-bnd1}),
(\ref{z-bnd2}), and (\ref{w2-minus}) imply
    \be
    y_+(0)=y_-(0)=C(\lambda_-)m(\lambda_-),~~~~~~~
    y_+'(0)=-e^{-4i\theta_\nu}y'_-(0)=e^{-4i\theta_\nu}C(\lambda_-),
    \label{bnd-minus}
    \ee
and consequently
    \be
    y_+(0)-e^{4i\theta_\nu}m(\lambda_-)y_+'(0)=0.
    \label{bnd-plus}
    \ee
In view of (\ref{z-E-pm}), which implies
$\lambda_+=e^{-4i\theta_\nu}\lambda_-$, and (\ref{mm=}) we can
express (\ref{bnd-plus}) as
    \be
    y_+(0)+\chi(\lambda_+)y_+'(0)=0,
    \label{chi}
    \ee
where
    \be
    \chi(\lambda):=\sum_{k=1}^\infty
    \frac{\sigma_k}{\lambda-e^{-4i\theta_\nu}\lambda_k},
    \label{chi=}
    \ee
Next, consider a fixed $\lambda_+\in\R$. Then because we have the
limit point case there is at most one linearly independent
$L^2$-solution $y_+$ of
    \be
    -y_+''(\rx)+\rx^{\nu+2}y_+(\rx)=\lambda_+ y_+(\rx).
    \label{y-plus=}
    \ee
This implies that $y_+^*$, which also belongs to $L^2[0,\infty)$
and solves (\ref{y-plus=}), satisfies
$y_+(\rx)^*=e^{i\gamma}y_+(\rx)$ for some $\gamma\in[0,2\pi)$.
Inserting this equation in the one obtained by taking the
complex-conjugate of both sides of (\ref{chi}) and using
$y_+(0)\neq 0\neq y_+'(0)$, we have
$\chi(\lambda_+)^*=\chi(\lambda_+)$. In view of (\ref{chi=}), the
latter relation reads $\Phi_1(\lambda_+)=0$ where
    \be
    \Phi_1(\lambda):=\sum_{k=1}^\infty
    \left(\frac{1}{\lambda-e^{4i\theta_\nu}\lambda_k}-
    \frac{1}{\lambda-e^{-4i\theta_\nu}\lambda_k}\right)\sigma_k.
    \label{phi-1=}
    \ee
Hence $\lambda_+$ is a real zero of $\Phi_1$. Clearly $\Phi_1$ is
a holomorphic function in $\C-{\cal S}_1^-\cup{\cal S}_1^+$ where
${\cal S}_1^\pm:=\{e^{\pm 4i\theta_\nu}\lambda_k\,|\, k\in\Z^+\}$.
Therefore, its  zeros (if exist) form a discrete set. This in turn
means that $\lambda_+$ and consequently the eigenvalues
$E_n=e^{2i\theta_\nu}\lambda_+$ (associated with this case, if
there are any) belong to discrete sets. The same argument applies
for the case $\lambda_-\in\R$. In summary, the eigenvalues that
lie on the rays: ${\rm arg}(z)=\pm 2i\theta_\nu$ form a possibly
empty discrete subset of $\C$. Next, we show that the same holds
for the eigenvalues lying outside these rays.

\item $\lambda_+\notin\R$ and $\lambda_-\notin\R$: In this case we
can use (\ref{w2}) to express $y_\pm$ as
    \be
    y_\pm(\rx)=C(\lambda_\pm)[y_1(\rx;\lambda_\pm)+m(\lambda_\pm)
    y_2(\rx;\lambda_\pm)].
    \label{w2=}
    \ee
Substituting this relation in (\ref{z-bnd1}) and (\ref{z-bnd2}),
we obtain
    \[C(\lambda_+)m(\lambda_+)=C(\lambda_-)m(\lambda_-),~~~~~~~
    C(\lambda_-)=-e^{4i\theta_\nu}C(\lambda_+).\]
These together with (\ref{z-E-pm}), (\ref{mm=}), and
$C(\lambda_\pm)\neq 0$ yield
    \be
    \Phi_2(E_n)=e^{2i\theta_\nu}m(e^{2i\theta_\nu}E_n)+
    e^{-2i\theta_\nu}m(e^{-2i\theta_\nu}E_n)=0,
    \label{eqn}
    \ee
where
    \be
    \Phi_2(\lambda):=-\sum_{k=1}^\infty
    \left(\frac{1}{\lambda-e^{2i\theta_\nu}\lambda_k}+
    \frac{1}{\lambda-e^{-2i\theta_\nu}\lambda_k}\right)\sigma_k.
    \label{w10}
    \ee
Therefore the eigenvalues $E_n$ are the zeros of
$\Phi_2$.\footnote{Note that in light of (\ref{m=}) we have
$\Phi_2(\lambda^*)=\Phi_2(\lambda)^*=\Phi_2(\lambda)$. Hence the
complex-conjugate of every zero of $\Phi_2$ is also a zero of
$\Phi_2$. This is consistent with the fact that the eigenvalues of
$H$ are either real or come in complex-conjugate pairs
\cite{p1,bender-98,bender-99}.} Clearly, $\Phi_2$ is a holomorphic
function in $\C-({\cal S}^-_2\cup {\cal S}^+_2)$ where ${\cal
S}_2^\pm:=\{e^{\pm 2i\theta_\nu}\lambda_k|k\in\Z^+\}$. This
implies that the zeros $E_n$ of $\Phi_2$ form a discrete set.
Hence the eigenvalues that do not lie on the rays ${\rm
arg}(z)=\pm 2i\theta_\nu$ also form a discrete set.
\end{enumerate}
This completes the proof that the set of all the eigenvalues $E_n$
is discrete.~~~$\square$
\end{itemize}

\newpage


\ed